\documentclass[conference]{IEEEtran}
\usepackage{cite,graphicx,amssymb,amsmath,color,textcomp,epsfig}
\usepackage{extarrows,multirow,multicol}
\usepackage{mathabx} 
\usepackage{bm}
\usepackage{booktabs}
\usepackage{subeqnarray}
\usepackage{subfig}
\usepackage{float}
\usepackage{enumerate}
\usepackage{algorithm}
\usepackage{multicol}
\usepackage{algpseudocode}

\newcommand{\erf}{\ensuremath{{\sf erf}}}
\newcommand{\erfc}{\ensuremath{{\sf erfc}}}

\newcommand{\Cov}{\mathrm{Cov}}
\usepackage{caption}
\newtheorem{remark}{\underline{Remark}}

\newlength{\figwidth}
\IEEEoverridecommandlockouts

\begin{document}
\title{
 Dynamic Scheduling for Enhanced Performance in RIS-assisted Cooperative Network with Interference
\vspace{-0.4cm}}
\author{
 \IEEEauthorblockN{Yomali Lokugama\IEEEauthorrefmark{1}, Saman~Atapattu\IEEEauthorrefmark{1}, Nathan Ross\IEEEauthorrefmark{2}, Sithamparanathan Kandeepan \IEEEauthorrefmark{1}, and Chintha Tellambura\IEEEauthorrefmark{3}}
 \IEEEauthorblockA{
\IEEEauthorrefmark{1}Department of Electrical and Electronic Engineering, School of Engineering, RMIT University, Victoria, Australia\\
\IEEEauthorrefmark{2}School of Mathematics and Statistics, University of Melbourne, Victoria, Australia\\\IEEEauthorrefmark{3}Department of Electrical Engineering, University of Alberta, Edmonton, Canada
\IEEEauthorblockA{Email:\IEEEauthorrefmark{1}\{yomali.lokugama,saman.atapattu\}@rmit.edu.au;\,\IEEEauthorrefmark{2}{nathan.ross}@unimelb.edu.au;\,\IEEEauthorrefmark{3}ct4@ualberta.ca}
}
}
\maketitle
\begin{abstract}
Reconfigurable Intelligent Surfaces (RIS) have emerged as transformative technologies, enhancing spectral efficiency and improving interference management in multi-user cooperative communications. This paper investigates the integration of RIS with Flexible-Duplex (FlexD) communication, featuring dynamic scheduling capabilities, to mitigate unintended external interference in multi-user wireless networks. By leveraging the reconfigurability of RIS and dynamic scheduling, we propose a user-pair selection scheme to maximize system throughput when full channel state information (CSI) of interference is unavailable. We develop a mathematical framework to evaluate the throughput outage probability when RIS introduces spatial correlation. The derived analytical results are used for asymptotic analysis, providing insights into dynamic user scheduling under interference based on statistical channel knowledge. Finally, we compare FlexD with traditional Full Duplex (FD) and Half Duplex (HD) systems against RIS-assisted FlexD. Our results show FlexD’s superior throughput enhancement, energy efficiency and data management capability in interference-affected networks, typical in current and next-generation cooperative wireless applications like cellular and vehicular communications.
\end{abstract}
\begin{IEEEkeywords}
Cooperative Networks, Dynamic Scheduling, Flexible Duplexing (FlexD), Interference Management, RIS. 
\end{IEEEkeywords}
\vspace{-0.2cm}
\section{Introduction}
Cooperative multiuser networks, which leverage Multiple-Input Multiple-Output (MIMO) systems, are essential for enhancing spectral efficiency in modern wireless communications. These networks enable resource sharing and virtual MIMO formation, benefiting applications like cellular and vehicular communications. The integration of Reconfigurable Intelligent Surfaces (RIS) represents a transformative advancement~\cite{Atapattu_ris_trans}, dynamically controlling signal propagation to improve signal strength and mitigate interference. This capability addresses critical challenges in cooperative networks, positioning RIS as a key enabler of next-generation wireless. Half-Duplex (HD) offers simplicity but low spectral efficiency~\cite{Ding}, while  Full-Duplex (FD) improves throughput theoretically double the rate but faces residual self-interference. Both lack adaptability in interference-heavy or insecure settings. FlexD~\cite{flexdPerera}, a dynamic scheduling framework, addresses these limitations by opportunistically selecting the direction based on channel and buffer conditions, thereby enhancing interference resilience and resource efficiency. Despite extensive studies on interference mitigation in HD and FD systems, the integration of FlexD-like scheduling in RIS-assisted multi-user interference networks remains largely unexplored. 

While interference mitigation in HD and FD networks is well studied~\cite{Askar_ieee,Shende,Goyal}, we focus on RIS-assisted multi-user interference networks. 
For \underline{HD-RIS} scenarios in interference networks, studies such as~\cite{Chen_Yali,Li_Zhengfeng,Ma_Yanan} have been conducted. \cite{Chen_Yali} investigates RIS-aided Device-to-Device (D2D) communications, enhancing link reliability. In~\cite{Li_Zhengfeng}, beamforming and RIS phase shifts are jointly optimized to maximize the weighted-sum rate. Similarly, \cite{Ma_Yanan} addresses the limitations of passive RISfor interference management by proposing a dual-functional active RIS. 
Similarly, \underline{FD-RIS} networks under interference have been widely studied, e.g.,~\cite{Karim_globecom,pala_wcnc,Ku_Chia_PIMRC,Tewes_Simon}. \cite{Karim_globecom} investigates STAR-RIS to improve coverage and spectral efficiency. \cite{pala_wcnc} investigates RIS-aided Vehicle-to-Everything (V2X) networks to maximise the achievable sum rate. Authors in ~\cite{Ku_Chia_PIMRC} introduce an interference cancellation scheme which maximises the system sumrate. 

However, the studies in~\cite{Chen_Yali,Li_Zhengfeng,Ma_Yanan,Karim_globecom,pala_wcnc,Ku_Chia_PIMRC,Tewes_Simon}, along with the references therein, predominantly employ fixed duplexing modes and do not incorporate dynamic scheduling. FlexD, which adaptively selects the transmission direction based on channel and interference conditions, remains unexplored in RIS-assisted networks. Even in non-RIS settings, FlexD-like approaches have primarily been studied for physical-layer security~\cite{Perera_grapgh}, with broader performance evaluations still lacking. 

\emph{This work is the first to explore the integration of RIS with FlexD for interference-aware dynamic scheduling.}. Key challenges include designing unified, adaptive frameworks that combine RIS control and FlexD scheduling for dynamic interference scenarios. Our key contributions are: i) \textit{Multi-User Network with External Interference}: We examine a multi-user network subject to external interference and propose a user-selection strategy that maximizes throughput while mitigating inter-user interference through orthogonal transmission. Outage probabilities are derived under partial channel state information (CSI) for the interference node, highlighting FlexD's superiority over static HD and FD in interference-prone scenarios; and ii)  \textit{Validation and Insights}: Simulations validate the proposed framework, demonstrating the enhanced performance of FlexD scheduling in mitigating interference compared to HD and FD. Additionally, a data availability model is integrated across the three schemes, emphasizing FlexD's role in improving energy efficiency.

\begin{figure*}
    \centering
    \subfloat[RIS-assisted network model with multiple users.\label{fig:network_model_pair}]{\includegraphics[width=0.33\textwidth]{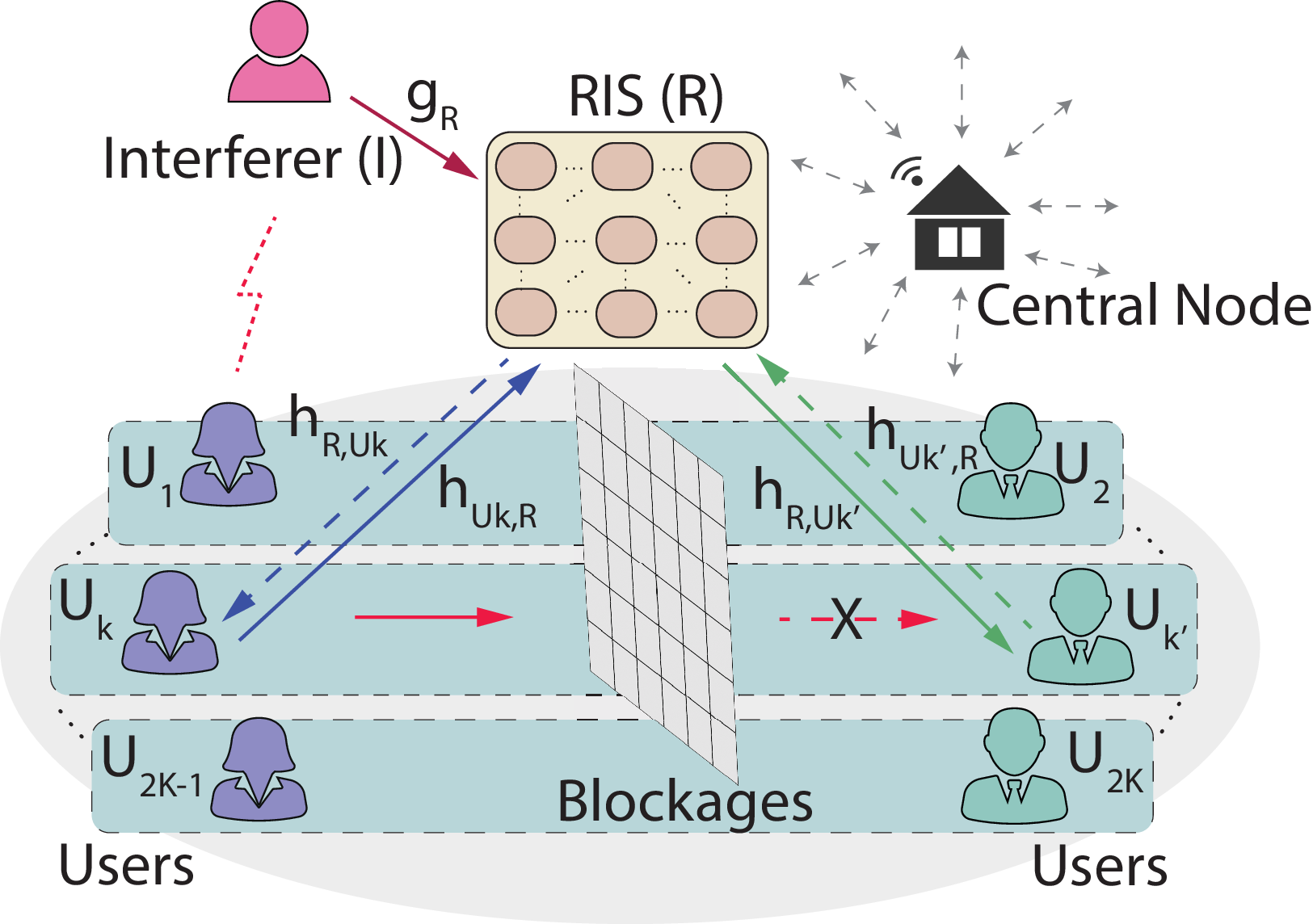}} 
    \hfill
     \subfloat[RIS-assisted VANET in an urban environment.\label{fig:network_model}]{\includegraphics[width=0.33\textwidth]{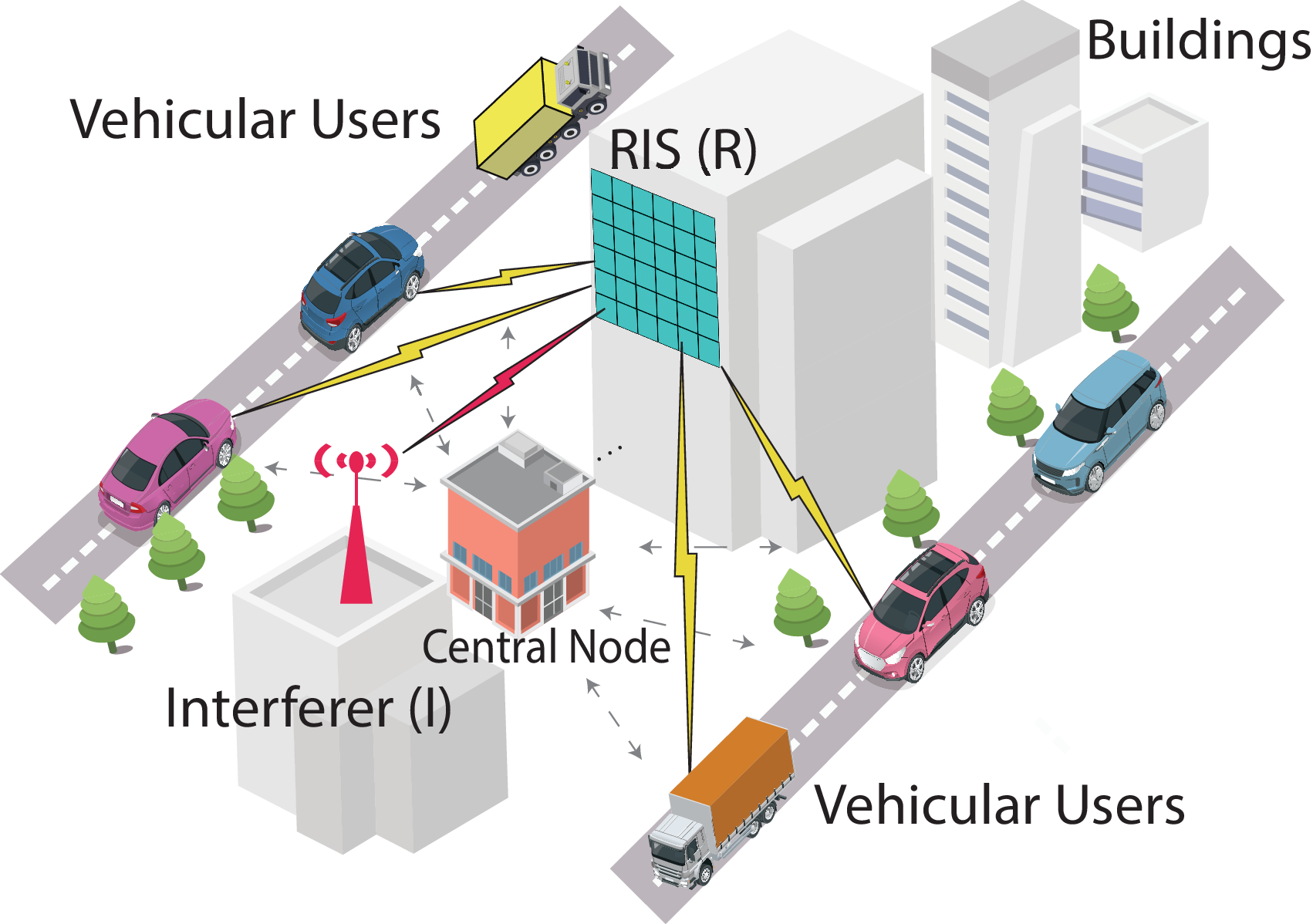}} \hfill
    \subfloat[Outage comparison of FlexD, HD, and FD. \label{fig:comparion_pair}]{\includegraphics[width=0.33\textwidth]{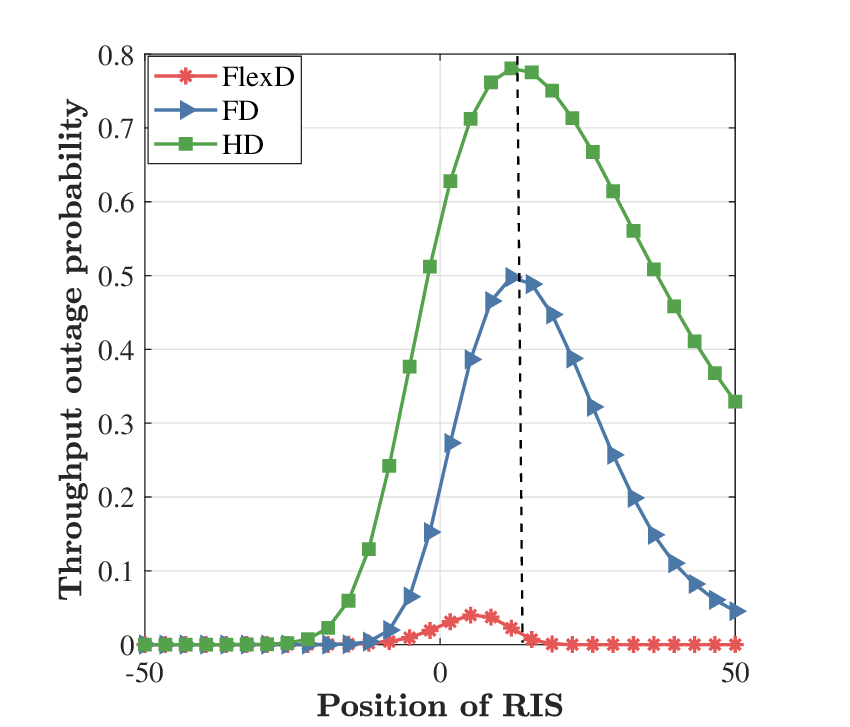}} 
    \caption{Illustration of the RIS-assisted network architecture and performance analysis. (a) Network model, (b) RIS-assisted VANET in an urban environment (c) Throughput outage comparison of FlexD, HD, and FD as a function of RIS position.}\vspace{-4.5mm}
\end{figure*}

\section{System Model}
\vspace{-2mm}
\subsection{Network Model}
\vspace{-1.5mm}
As illustrated in Fig.~{\ref{fig:network_model_pair}}, the network consists of \( 2K \) single-antenna communication users, denoted as \( \textsf{U}_1, \textsf{U}_2, \ldots, \textsf{U}_{2K} \). Direct communication between users is not feasible due to blockages and severe path loss. Consequently, user communication occurs via a RIS (\( \textsf{R} \)), which consists of $M$ reconfigurable elements. Additionally, the network includes a common single-antenna interference source, denoted as \( \textsf{I} \). \( \textsf{I} \) can be a transmitter from a neighbouring network sharing the same resource block or an intentional jammer.  \( \textsf{I} \) has weak direct links to the users but imposes significant interference on \( \textsf{R} \). Such a network is a typical example of a vehicular ad-hoc network (VANET) in urban environments, as shown in Fig.{\ref{fig:network_model}}. 
Users are grouped into \( K \) adjacent pairs. Each pair \((\textsf{U}_k, \textsf{U}_{k'})\), where \( k \in \{1, 2, \ldots, 2K\} \) and \( k' = 2(k \mod 2) + k - 1 \), shares a resource block. Communication utilizes FlexD mode via the RIS (\( \textsf{R} \)), enabling the optimal user pair (selected as per Section~\ref{sec_user_sel}) to dynamically alternate transmission and reception roles within the same time-frequency resource block. 
The small-scale fading coefficients between $\textsf{U}_{k}\rightarrow \textsf{R}$, $\textsf{R}\rightarrow \textsf{U}_{k'}$ and $\textsf{I}\rightarrow \textsf{R}$ are defined as   
$\bm{h}_{\textsf{U}_k,\textsf{R}} = \{h_{k,m}\}_{m=1}^M \in \mathbb{C}^{M \times 1}$, 
$\bm{h}_{\textsf{R},\textsf{U}_{k'}} = \{h_{m,k'}\}_{m=1}^M \in \mathbb{C}^{M \times 1}$, and
$\bm{g}_{\textsf{R}} = \{g_{m}\}_{m=1}^M \in \mathbb{C}^{M \times 1}$, respectively. 
Moreover, the fading between \(\textsf{U}_k \to \textsf{R}\) is modeled as
$\bm{h}_{\textsf{U}_k,\textsf{R}} \sim \mathcal{CN}\left(\mathbf{0}, A\zeta\mathbf{\Omega}\right)$, 
where \( A \) is the area of each RIS element, \( \zeta \) is the average intensity attenuation (i.e, $\sigma^2=A\zeta$), and \( \mathbf{\Omega} = \{\Omega_{i,j}\}_{i,j=1}^M \in \mathbb{C}^{M \times M} \) is the normalized spatial correlation matrix. 
The correlation matrix elements are given by  
$
\left[ \mathbf{\Omega} \right]_{i,j} = \mathrm{sinc}\left( \frac{2\pi \| \mathbf{a}_i - \mathbf{a}_j \|}{\lambda_w} \right),
$
where \( \lambda_w \) is the wavelength, and \( \| \mathbf{a}_i - \mathbf{a}_j \| \) denotes the distance between elements \( i \) and \( j \) ~\cite{coon_com_let_2022}. Similarly, the fading channels
$\bm{h}_{\textsf{R},\textsf{U}_{k'}} \sim \mathcal{CN}\left(\mathbf{0}, A\zeta\mathbf{\Omega}\right)$ and 
$\bm{g}_{\textsf{R}} \sim \mathcal{CN}\left(\mathbf{0}, A\zeta\mathbf{\Omega}\right)$. 
\vspace{-1mm}
\subsection{Signal Model}
For the reflection mode, the incident signals are reflected by the  RIS elements without power loss and the reflecting coefficients matrix can be given as $\mathbf{\Psi}=\text{diag}(e^{j\psi_1},\ldots,e^{j\psi_{M}}) \in \mathbb{C}^{M \times M}$. Then, the received signal at $\textsf{U}_{k'}$ when $\textsf{U}_k$ transmits can be given as 
\begin{align}
    y_{k,k'}(t)= & \sqrt{P/d_{k,\textsf{R}}^\alpha d_{\textsf{R},k'}^\alpha} 
(\bm{h}_{\textsf{U}_k,\textsf{R}}^T \mathbf{\Psi} \bm{h}_{\textsf{R},\textsf{U}_{k'}}) s_k(t) 
\notag\\&+ \sqrt{Q/d_{\textsf{R},k'}^\alpha d_{\textsf{I},\textsf{R}}^\alpha}
(\bm{h}_{\textsf{R},\textsf{U}_{k'}}^T \mathbf{\Psi} \bm{g}_{\textsf{R}}) s_{\textsf{I}}(t) 
+ w_{k'}(t)
\end{align}
same as at $\textsf{U}_{k}$ when $\textsf{U}_{k'}$ transmits
\begin{align}
    y_{k',k}(t)= & \sqrt{P/d_{k',\textsf{R}}^\alpha d_{\textsf{R},k}^\alpha} 
(\bm{h}_{\textsf{U}_{k'},\textsf{R}}^T \mathbf{\Psi} \bm{h}_{\textsf{R},\textsf{U}_{k}}) s_{k'}(t) 
\notag\\&+ \sqrt{Q/d_{\textsf{R},k}^\alpha d_{\textsf{I},\textsf{R}}^\alpha}
(\bm{h}_{\textsf{R},\textsf{U}_{k}}^T \mathbf{\Psi} \bm{g}_{\textsf{R}}) s_{\textsf{I}}(t) 
+ w_{k}(t)
\end{align}
where $P$ and $Q$ are the transmit powers of the active users and the interference node, respectively, while $d_{k,\textsf{R}}$, $d_{\textsf{R},k'}$, and $d_{\textsf{I},\textsf{R}}$ denote the corresponding distances. The path-loss exponent is $\alpha$. The transmitted signals of $\textsf{U}_k$ and $\textsf{I}$ are $s_k$ and $s_{\textsf{I}}$, respectively, having unit-power, i.e., $\mathbb{E}[|s_k(t)|^2] = 1$. The term $w_{k'}$ represents additive white Gaussian noise (AWGN) with $ w_{k'} \sim \mathcal{CN}(0, \sigma_n^2)$.

We assume full instantaneous CSI for the user channels, \(\bm{h}_{\textsf{U}_k,\textsf{R}}\) and \(\bm{h}_{\textsf{R},\textsf{U}_{k'}}\), \(\forall k, k'\), while only statistical information, such as \(\mathbb{E}[\bm{g}_{\textsf{R}}]\) and \(\mathrm{Var}[\bm{g}_{\textsf{R}}]\), is available for the interferer or jammer channels. This assumption is realistic, as instantaneous estimation of unintended user channels is challenging. Consequently, RIS phase shifts are optimized in real-time based solely on \(\bm{h}_{\textsf{U}_k,\textsf{R}}\) and \(\bm{h}_{\textsf{R},\textsf{U}_{k'}}\) to maximize the received power of the desired user signal.

The channel coefficients are expressed as \(h_{k,m} = \alpha_{k,m} e^{-j\theta_{k,m}}, h_{m,k'} = \alpha_{m,k'} e^{-j\theta_{m,k'}}\), and \(g_m = \beta_m e^{-j\phi_m}\), $\forall k, k',m$. For \(\textsf{U}_k \to \textsf{U}_{k'}\) or \(\textsf{U}_{k'} \to \textsf{U}_k\) communication, the \(m\)th RIS phase adjustment is given by \(\psi_m = \theta_{k,m} + \theta_{m,k'}\) or \(\psi_m = \theta_{k',m} + \theta_{m,k}\), respectively. The resulting Signal-to-Interference-plus-Noise Ratio (SINR)s are then derived as

\begin{equation} 
    \gamma_{k,k'} = \frac{ \bar{\gamma}_{k,k'}\left(A_{k,k'} \right)^2}{\eta_{I,k'}\left|{B_{I,k'}}\right|^2\hspace{-1mm} +1} \text{ and }
    \gamma_{k',k} = \frac{ \bar{\gamma}_{k',k}\left(A_{k',k} \right)^2}{\eta_{I,k}\left|{B_{I,k}}\right|^2\hspace{-1mm} +1},
    \label{eq:sinr}
\end{equation}
respectively, where $\bar{\gamma}_{k,k'}=\bar{\gamma}_{k',k}=P/d_{k,\textsf{R}}^\alpha d_{\textsf{R},k'}^\alpha \sigma_n^2 , \eta_{I,k'}=Q/d_{\textsf{I},\textsf{R}}^\alpha d_{\textsf{R},k'}^\alpha \sigma_n^2$ and $\eta_{I,k}=Q/d_{\textsf{I},\textsf{R}}^\alpha d_{\textsf{R},k}^\alpha \sigma_n^2$. Here, random variables $A_{k,k'}$ and $A_{k',k}$ are real-valued scalars, while $B_{I,k'}$ and $B_{I,k}$ are complex values, given as 
\begin{align*}
&A_{k,k'}=\hspace{-1.5mm}\sum_{m=1}^{M}\alpha_{k,m}{\alpha_{m,k'}} ,{B_{I,k'}}= \hspace{-2mm}\sum_{m=1}^{M}\beta_{m}{\alpha_{m,k'}} e^{-j(\phi_{m}-\theta_{k,m})},  \\&A_{k',k}=\hspace{-1.5mm}\sum_{m=1}^{M}\alpha_{k',m}{\alpha_{m,k}},{B_{I,k}}=\hspace{-1.5mm} \sum_{m=1}^{M}\beta_{m}{\alpha_{m,k}} e^{-j(\phi_{m}-\theta_{k',m})}
\end{align*}   

\subsection{FlexD Motivation in a Single-User Pair Scenario}  
As this is the first study to analyze FlexD under interference, we begin by motivating its design through a comparative analysis with conventional HD and FD modes in a simple network comprising a single user pair and one interferer.  Consider a single user pair, \(\textsf{U}_k\) and \(\textsf{U}_{k'}\), assisted by a RIS with \(M = 128\) elements. The nodes are located on a 2D plane at \(\textsf{U}_k(-30, 40)\), \(\textsf{U}_{k'}(50, -30)\), and \(\textsf{I}(10, 15)\). RIS \((\textsf{R}\)) moves linearly along the x-axis from \((-50, 0)\) to \((50, 0)\). In HD mode, communication is unidirectional from \(\textsf{U}_k\) to \(\textsf{U}_{k'}\); FD enables simultaneous bidirectional transmission. FlexD dynamically selects the transmission direction with the higher SINR. 

Fig.~\ref{fig:comparion_pair} shows the simulated throughput outage probabilities of HD, FD, and FlexD versus the position of the RIS \(\textsf{R}\) under the described setup. FlexD consistently outperforms {\it HD, which uses a fixed communication direction regardless of RIS position}. It also outperforms {\it FD, which, despite utilizing full resources for simultaneous bidirectional transmission, neither adapts to interference dynamics nor mitigates residual self-interference (RSI) effectively}. In contrast, FlexD opportunistically selects the more reliable direction based on the interference impact, enabling superior resilience with digital signal processing (DSP) complexity comparable to HD, whereas FD demands advanced DSP techniques for self-interference suppression. These results highlight FlexD’s potential for efficient operation in more complex multi-user scenarios, which will be investigated in subsequent sections.

\subsection{User Selection in FlexD Network}\label{sec_user_sel}

For the user pair \((\textsf{U}_k, \textsf{U}_{k'})\), the communication direction for FlexD \(d_{k \leftrightarrow k'}\) is selected to maximize the user SINR in \eqref{eq:sinr}. In this centralized model, a central node manages CSI for all
nodes to optimize direction control. Thus, the communication direction and the respective SINR can be expressed as
\begin{align}
    (d_{k \leftrightarrow k'}, \gamma_{k \leftrightarrow k'}) = 
\begin{cases}
  (k \rightarrow k', {\gamma}_{k,k'}) & \text{if } {\gamma}_{k,k'} \geq {\gamma}_{k',k}, \\
  (k' \rightarrow k, {\gamma}_{k',k}) & \text{if } {\gamma}_{k,k'} < {\gamma}_{k',k}.
\end{cases}
\end{align}
However, the decision node cannot directly compute \(\gamma_{k,k'} \gtreqless \gamma_{k',k}\) due to the unavailability of \(\beta_m\) and \(\phi_m\). Hence, we replace \(B_{I,k'}\) and \(B_{I,k}\) in \eqref{eq:sinr} with their expected values with respect to $\bm{g}_{\textsf{R}}$, i.e., \(\mathbb{E}_{\bm{g}_{\textsf{R}}}[B_{I,k'}]\) and \(\mathbb{E}_{\bm{g}_{\textsf{R}}}[B_{I,k}]\), and perform \(\hat{\gamma}_{k,k'} \gtreqless \hat{\gamma}_{k',k}\) instead, where \(\hat{\gamma}_{k,k'}=   \bar{\gamma}_{k,k'}(A_{k,k'})^2/(\eta_{I,k'} \sigma ^2(\sum_{m=1}^{M}{\alpha_{m,k'}^2}) +1)\) and \(\hat{\gamma}_{k',k}=\bar{\gamma}_{k',k}(A_{k',k})^2/(\eta_{I,k} \sigma ^2(\sum_{m=1}^{M}{\alpha_{m,k}^2}) +1)\). The effective communication direction and SINR for the user pair \((\textsf{U}_k, \textsf{U}_{k'})\) is then given by
\begin{align}
\label{eq_sinr_pcsi}
    (d_{k \leftrightarrow k'}, \gamma_{k \leftrightarrow k'}) = 
\begin{cases}
  (k \rightarrow k', {\gamma}_{k,k'}) & \text{if } \hat{\gamma}_{k,k'} \geq \hat{\gamma}_{k',k}, \\
  (k' \rightarrow k, {\gamma}_{k',k}) & \text{if } \hat{\gamma}_{k,k'} < \hat{\gamma}_{k',k}.
\end{cases}
\end{align}
The remaining analysis will be based on the user selection criteria in a multi-user network with the effective SINR of user pair \((\textsf{U}_k, \textsf{U}_{k'})\) described in \eqref{eq_sinr_pcsi}.
A user selection scheme maximizes throughput across all pairs, is given as
\begin{align}\label{eq_R}
    {R} = \max_{k \in S} \, R_{k \leftrightarrow k'}, \quad R_{k\leftrightarrow k'} = \ln(1+\gamma_{k\leftrightarrow k'})/2
\end{align}
where $S = \{1, 3, 5, \dots, 2K-1\}$ represents odd-numbered users and $R_{k\leftrightarrow k'}$ represents throughput for a user pair $\textsf{U}_k-\textsf{U}_{k'}$ in FlexD network for SINR described in \eqref{eq_sinr_pcsi}. 
\section{Performance Analysis}
Performance is evaluated using the throughput outage probability, measuring the likelihood that system throughput falls below a threshold \( R_t \). This metric provides insight into the system's reliability. The next section rigorously derives the outage probability with an asymptotic analysis for additional insights. The throughput outage probability is evaluated as 
\begin{align}
     P_{\text{o}} &\stackrel{(a)}{=} \Pr\left(\max_{k \in S} R_{{k  \leftrightarrow k'}} \leq R_{\text{t}}\right) 
     \stackrel{(b)}{=} \prod_{k=1}^{K} \Pr\left(R_{{k  \leftrightarrow k'}} \leq R_{\text{t}}\right) \notag\\
     & \stackrel{(c)}{=} \prod_{k=1}^{K} \Pr\left(\frac{1}{2} \ln(1+\gamma_{k\leftrightarrow k'}) \leq R_{\text{t}}\hspace{-0.5mm}\right)\hspace{-1mm}
     \stackrel{(d)}{=}\hspace{-1mm} \prod_{k=1}^{K} F_{\gamma_{{k  \leftrightarrow k'}}}\left(\tau\right) 
     \label{eq_pout_general}
\end{align}
where $\tau = e^{2R_{\text{t}}} - 1$. The derivation steps are as (a) is based on the outage definition; (b) holds due to the independence of channels and interference across users, leading to independent rates $R_{{k \leftrightarrow k'}}$ (i.e., even though $\bm{g}_{\textsf{R}}$ is common for every user pair in the network, we can show that $\gamma_{k \leftrightarrow k'}$ and $\gamma_{\ell \leftrightarrow \ell'}$ are uncorrelated to each other, where  $\gamma_{k \leftrightarrow k'}$ and $\gamma_{\ell \leftrightarrow \ell'}$ are the SINR for two different user pairs \((\textsf{U}_k, \textsf{U}_{k'})\) and \((\textsf{U}_\ell, \textsf{U}_{\ell'})\) by showing the covariance between $B_{I,k'}$ and ${B_{I,\ell'}}$. 
($\Cov\left({B_{I,k'}},{B_{I,\ell'}}\right) =0$)); (c) derives from the relationship between rate and SINR; and (d) expresses the probability in terms of the cumulative distribution function (CDF) of $\gamma_{{k \leftrightarrow k'}}$ from \eqref{eq_sinr_pcsi}, with $F_{\gamma_{{k \leftrightarrow k'}}}\left(\tau\right)$ denoting the CDF of SINR. 
Once the CDF $F_{\gamma_{{k \leftrightarrow k'}}}\left(\tau\right)$ is established for various scenarios, the outage probability  $P_{\text{o}}$ can be readily evaluated. 
The CDF $F_{\gamma_{{k \leftrightarrow k'}}}\left(\tau\right)$ can be evaluated as
\begin{align}
     F_{\gamma_{{k  \leftrightarrow k'}}}\left(\tau\right) = &\Pr\left(  \gamma_{k,k'} < \tau  \quad 
  \&  \quad
\hat{\gamma}_{k,k'}  \geq \hat{\gamma}_{k',k} \right)\label{outage}
 \\&+\Pr\left(  \gamma_{k',k} < \tau  \quad 
  \&  \quad
\hat{\gamma}_{k,k'}  < \hat{\gamma}_{k',k} \right).\notag
\end{align}
After the simplification, this can be approximated as 
\begin{align}
\label{outage:case1}
    F_{\gamma_{{k  \leftrightarrow k'}}}\left(\tau\right) \hspace{-1mm}\approx \hspace{-1mm}
\begin{cases}
 \mathcal{F}(\bar{\gamma}_{k,k'},\eta_{I,k'},s,\mu,\rho,\tau) & \text{if } \eta_{I,k} \geq \eta_{I,k'}, \\
  \mathcal{F}(\bar{\gamma}_{k',k},\eta_{I,k},s,\mu,\rho,\tau) & \text{if } \eta_{I,k} < \eta_{I,k'}.
\end{cases}
\end{align}
where
\begin{align}
&\mathcal{F}(\delta,\epsilon,s,\mu,\rho,\tau)  = 1+C\left( 1+ \erf\left(\frac{\left(\frac{\mu}{s}-D\right)}{G} \right)\right)
  \notag \\&+\hspace{-1mm}  C\left( \erfc\left(\hspace{-1mm}\frac{\left(\frac{\mu}{s}+D\right)}{G} \hspace{-1mm}\right)\hspace{-1mm}\right)-\frac{1}{2}\erfc\left(L \right)-\frac{1}{2}\erfc\left(N\right),\notag
\end{align}
\begin{align}
&\text{with } L=\frac{\sqrt{\tau}-\mu\sqrt{\delta}}{\sqrt{2s\delta}} ,N=\frac{\sqrt{\tau}+\mu\sqrt{\delta}}{\sqrt{2s\delta}},\notag
    \\&C=\frac{\sqrt{\tau \epsilon}}{2\sqrt{2s\rho \delta +\tau \epsilon}}e^{\left(\frac{\mu^2\tau \epsilon}{2s(2 s\rho \delta + \tau \epsilon)}+\frac{\rho}{\epsilon}- \frac{\mu^2}{2s}\right)},
    \notag\\&D=\left(\frac{2 \delta \rho s+\tau \epsilon}{s\tau \epsilon}\right)\sqrt{\frac{\tau}{\delta}}, G=2\sqrt{\frac{2 s \rho \delta +\tau \epsilon}{2s\tau \epsilon}}
    \notag
\end{align}
where $\sigma^2 = A\zeta, \mu = \pi \sigma^2 M/4 , s=\sum_{i=1}^{M} \sum_{j=1}^{M} \left (\sigma^2{\left [{\bar {\mathbf {\Omega}} }\right]_{i, j} }\right)^{2}\hspace{-2mm} - \pi ^{2}\sigma^4M ^{2}/16 ,\, \rho=1/(\sum_{i=1}^{M} \sum_{j=1}^{M} \sigma^4 \mathbf {\Omega}_{i,j}^2[\bar {\mathbf {\Omega}}]_{i,j}^2)$ and   
$\erf\left[\cdot\right]$ is  the Gauss error function and $\erfc\left[\cdot\right]$ is  the complementary Gauss error function. 
\begin{equation}
   \left [{\bar {\mathbf {\Omega}} }\right]_{i, j} = 
\begin{cases}
 \frac {\left |{\left [{\mathbf {\Omega} }\right]_{i, j} }\right |^{2} - 1}{2} {\sf K}\left ({{\left |{\left [{\mathbf {\Omega} }\right]_{i, j} }\right |^{2}}}\right) \hspace{-0.5mm}+\hspace{-0.5mm}  {\sf E}\left ({{\left |{\left [{\mathbf {\Omega}}\right]_{i, j} }\right |^{2}}}\right) \hspace{-2mm}&\hspace{-2mm}; i \neq j, \\
  1 & \hspace{-2mm}; i=j
\end{cases}\notag
\end{equation}
where ${\sf K}\left(\cdot\right)$ and ${\sf E}\left(\cdot\right)$ are the complete elliptic integral first and second kind, respectively. Proof is in Appendix. 

For the high-SINR regime, where \(P, Q \to \infty \, \forall k\), asymptotic behaviour of the network can be expressed as $P_{\text{o}}^\infty = \prod_{k=1}^{K}P_{\text{o}(k\leftrightarrow k')}^\infty$, where $P_{\text{o}(k\leftrightarrow k')}^\infty$ is the outage at high-SINR of user pair \((\textsf{U}_k, \textsf{U}_{k'})\). 
\begin{align}
\label{outage:asym}
   P_{\text{o}(k\leftrightarrow k')}^\infty \hspace{-1mm}= \hspace{-1mm}
\begin{cases}
 \mathcal{F}^\infty(d_{k,\textsf{R}}^\alpha,d_{\textsf{I},\textsf{R}}^\alpha,s,\mu,\rho,\tau) & \hspace{-2mm}\text{if } d_{\textsf{R},k'}^\alpha \geq d_{\textsf{R},k}^\alpha, \\
  \mathcal{F}^\infty(d_{k',\textsf{R}}^\alpha,d_{\textsf{I},\textsf{R}}^\alpha,s,\mu,\rho,\tau) & \hspace{-2mm}\text{if } d_{\textsf{R},k'}^\alpha < d_{\textsf{R},k}^\alpha.
\end{cases}
\end{align}
where
\begin{align*}
\mathcal{F}^\infty(\nu,d_{\textsf{I},\textsf{R}}^\alpha,s,\mu,\rho,\tau)\hspace{-1mm}=\hspace{-1mm}\widetilde{C}\left(\hspace{-1mm}1\hspace{-1mm}+\hspace{-1mm}\erf\left(\hspace{-1mm}\frac{\mu}{s \widetilde{G}}\hspace{-1mm}\right)\hspace{-1mm}\right)\hspace{-1mm}+\hspace{-1mm}\widetilde{C}\left(\hspace{-1mm}\erfc\hspace{-1mm}\left(\hspace{-1mm}\frac{\mu}{s \widetilde{G}}\right)\hspace{-1mm}\right)\hspace{-1mm}
\end{align*}
\normalsize
\begin{align}
\text{with } &\widetilde{C}=\frac{\sqrt{\tau \nu}}{2\sqrt{2s\rho d_{\textsf{I},\textsf{R}}^\alpha +\tau \nu}}e^{\left(\frac{\mu^2\tau \nu}{2s(2 s\rho d_{\textsf{I},\textsf{R}}^\alpha + \tau \nu)}- \frac{\mu^2}{2s}\right)},
    \notag\\&\widetilde{G}=2\sqrt{\left(2 s \rho d_{\textsf{I},\textsf{R}}^\alpha +\tau \nu\right)/{2s\tau \nu}}.
   \notag 
\end{align}
\begin{remark}
Since  $P_{\text{o}}^\infty$ depends mainly on distances, maintaining a significant distance between receiving nodes and $\textsf{I}$ is crucial (see Fig.~{\ref{fig:comparion_pair}}). 
This highlights FlexD's dynamic scheduling importance over fixed HD or FD for interference-aware networks.
\end{remark}

\vspace{-2mm}
\section{Numerical Analysis}
This section evaluates the analytical framework via simulations in a circular network area with a radius of \(1\)\,km with six user pairs (i.e., $K=6$) and one interference node. Users in each pair is located around \(100\)\,-\,\(600\)\,m from \(\textsf{R}\), while the interference node is positioned \(300\)\,m apart from \(\textsf{R}\). All wireless channels are modelled as $\sim \mathcal{CN}\left(\mathbf{0}, A\zeta\mathbf{\Omega}\right)$ with $A=d^2$, we consider square RIS elements each with width ($d_H$) and height ($d_V$), where $d=d_H=d_V$. $\mathbf{\Omega}$ is calculated with wavelength $\lambda_w=0.1\,$m. Unless or otherwise stated $d$ is set to $\lambda_w/8$. Path loss per unit distance between user pairs and the RIS is set to \(-30\)\,dB while interference node and RIS is set to \(-35\)\,dB with path loss exponent \(\alpha = 2.5\).
\begin{figure*}
    \centering
    \subfloat[Outage vs. $P_\text{max}$ for different $M$ values.\label{fig_all}]{\includegraphics[width=0.33\textwidth]{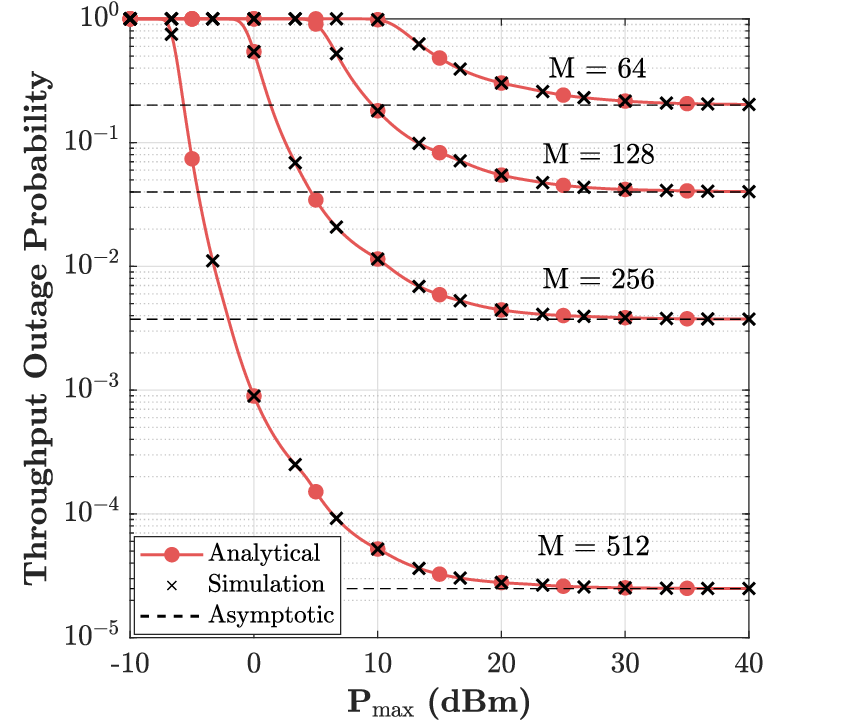}} 
    \hfill
    \subfloat[Outage vs. $\text{SINR}(\text{dB})$ for FlexD, HD and FD. \label{fig:comparison}]{\includegraphics[width=0.33\textwidth]{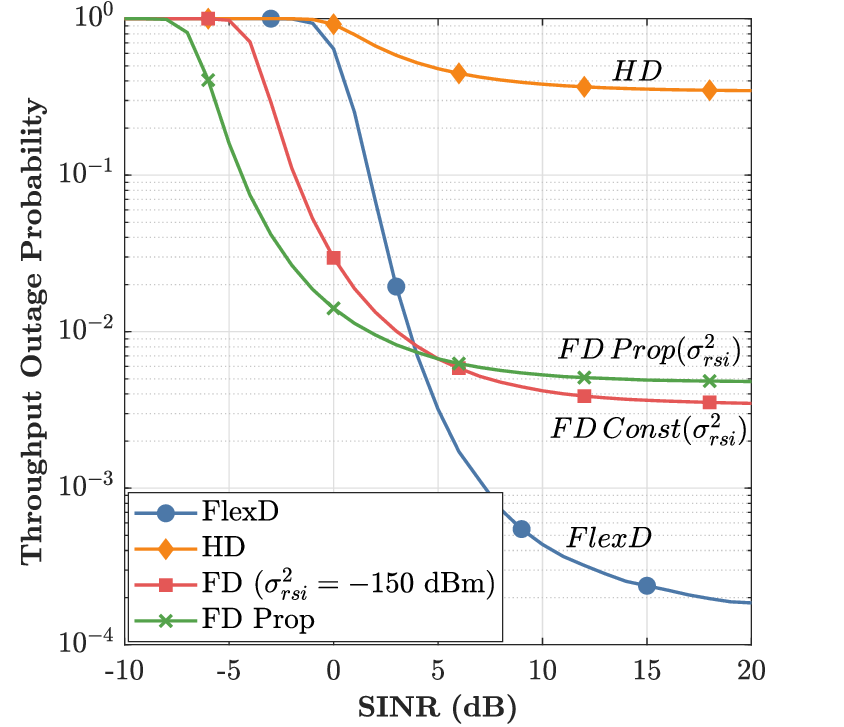}} \hfill
    \subfloat[EE vs. $P_\text{max}$ for FlexD, HD and FD.\label{fig_EE}]{\includegraphics[width=0.33\textwidth]{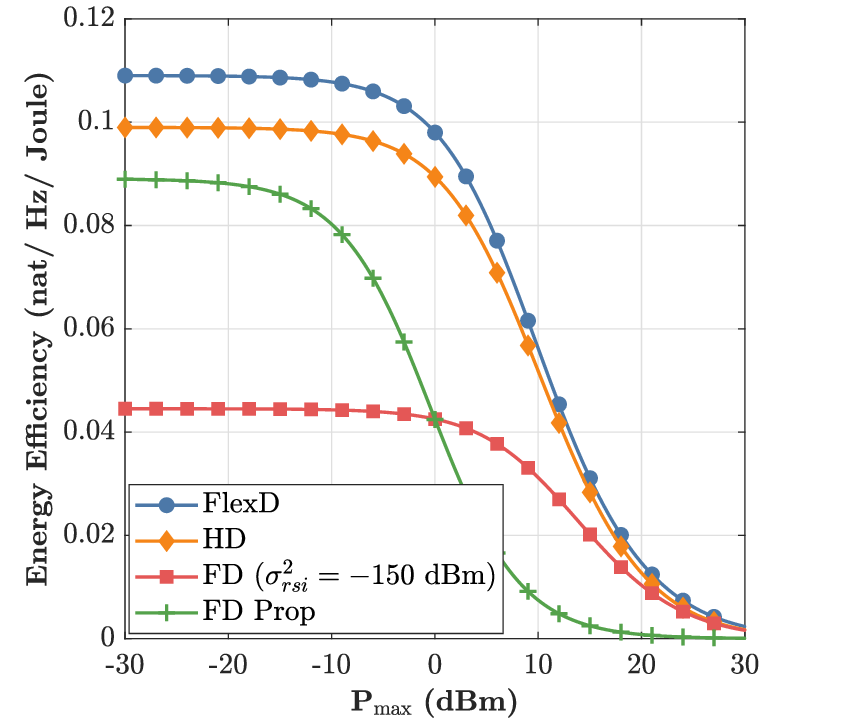}} 
    \caption{Performance analysis for different scenarios and FlexD, HD, and FD comparison with \(K=6\) and $d=\lambda_w/8$.}\vspace{-4.5mm}
\end{figure*}
\subsection{Validation and Performance Comparison}

Fig.~\ref{fig_all} shows the throughput outage probability versus power (\(P_\text{max}\)) for different values of \(M\) with RIS element dimension \(d=\lambda_w/8\). The close agreement confirms the accuracy of analytical and simulation results from \eqref{outage:case1} and \eqref{eq_sinr_pcsi}, respectively. As \(P_\text{max}\) increases, all scenarios reach an outage floor, where performance is limited by interference, as confirmed by the asymptotic analysis in \eqref{outage:asym}. As expected, when the number of RIS elements increases, throughput outage reduces due to enhanced spatial diversity. For example, with \(M=256\), performance improves by approximately \(98\%\) over \(M=64\) at \(10\) dBm. This demonstrates the accuracy between the simulation and analytical results, even with the smaller $d$ value, indicating a higher correlation between elements reduces the number of independent signal copies \cite{coon_com_let_2022}. 

Fig.~\ref{fig:comparison} compares the throughput outage probability versus SINR (dB) for the FlexD system, HD, and two-way FD~\cite{ZhangZhengquan2016} with \(M=128\) and \(d=\lambda_w/8\) with same network. In HD, communication occurs unidirectionally, while FD allows simultaneous bidirectional transmission. The respective throughputs for HD, two-way FD, and FlexD are given by
    $R_{\text{HD}} =  \ln(1+\gamma_{k,k'})/2$,  
    $R_{\text{FD}} = \min(\ln(1+\gamma_{k,k'}^{FD}),\ln(1+\gamma_{k',k}^{FD}))$~\cite[Eq.~(9)]{ZhangZhengquan2016} and 
    $R_{\text{FlexD}} = \ln(1+\gamma_{k\leftrightarrow k'})/2$,
where the $1/2$ factor in HD and FlexD accounts for half-duplex operation  and $\gamma_{k,k'}^{\text{FD}}$ and $\gamma_{k',k}^{\text{FD}}$ include residual self-interference (RSI) due to full-duplex operation. For FD, we evaluate two RSI schemes: one with constant RSI (\(\sigma_{rsi}^2\)) at \(-150\) dBm (the noise floor) and another that is linearly dependent on transmit power~\cite[Eq.~(8), with \(\lambda = 1\)]{Leonardo2014}. FD outperforms HD, due to simultaneous bidirectional transmissions, while FlexD achieves even better performance. Although FD utilizes all resources continuously, it lacks adaptive interference management. In contrast, FlexD dynamically manages resources, delivering superior user rates.

The energy efficiency (EE), measured in nat/Hz/Joule, is defined as the ratio of achievable throughput to total consumed power ($P_{total}$), expressed as \(R / P_{total}\). For HD and FlexD, the total consumed power is \(P_\text{max}\), while for FD, it is \(2P_\text{max}\), assuming full power allocation for a given time instance. In Fig.~\ref{fig_EE}, we compare FlexD network EE against maximum power \(P_\text{max}\) with equivalent HD and FD systems. FlexD consistently outperforms both HD and FD across the entire region of \(P_\text{max}\), Illustrating the importance of dynamic scheduling. For instance, at \(P_\text{max} = -20\)\,dBm, the EE values are \(0.108\), \(0.098\), and \(0.044\) [nat/Hz/Joule] for FlexD, HD, and FD with RSI at $-150$\,dBm, respectively.
\vspace{-0.2cm}

\begin{figure}[t!]
  \centering
\includegraphics[width=0.35\textwidth]{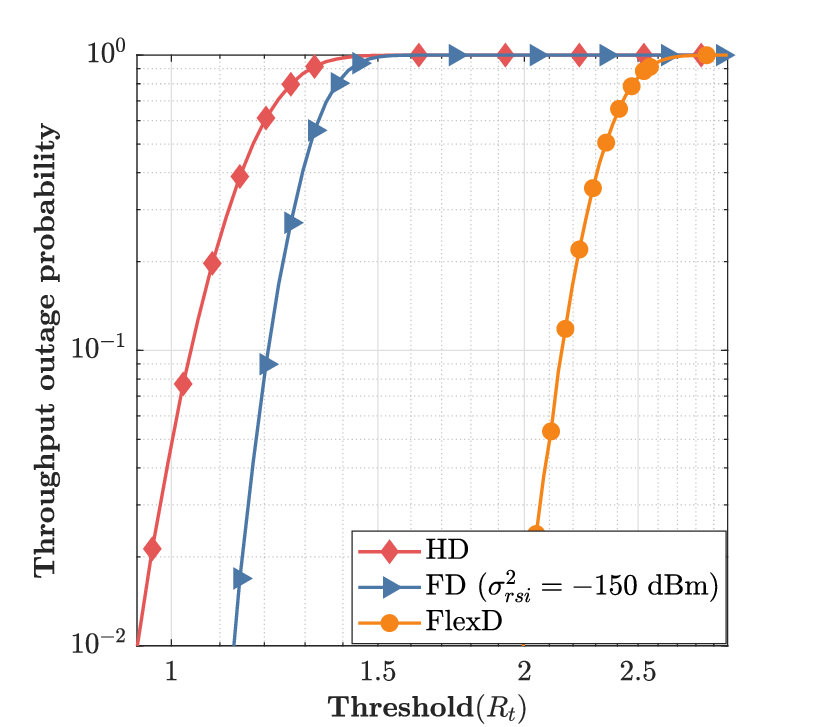}
  \caption{Performance under data availability.}\vspace{-4.5mm}
  \label{fig:que}
\end{figure}

\subsection{Traffic and Data Availability Models}
To realistically evaluate HD, FD, and FlexD in RIS-assisted systems, it is crucial to model both the wireless channel and user-side data availability, as throughput, particularly in FlexD, is closely tied to buffer states and traffic dynamics. This section defines the traffic models used to evaluate HD, FD, and FlexD in a RIS-assisted single user pair network \((\textsf{U}_k, \textsf{U}_{k'})\). Each user independently generates packets following a Poisson process \cite{MIT_OCW_Stochastic_2011} in each time slot \(t\) over a simulation duration \(T\), with arrivals queued in a First-In, First-Out (FIFO) buffer. Transmission scheduling depends on the duplexing mode and is governed by channel capacity and buffer occupancy. Table~\ref{table:1} summarizes each mode as 
\begin{itemize}
    \item \textbf{HD:} Only one user transmits per slot. If both users have data, one is selected at random. Otherwise, the user with data transmits. No simultaneous transmission is allowed.   
    \item \textbf{FD:} Both users can transmit and receive simultaneously. Each transmits up to \(\min(C_k, B_k)\) and \(\min(C_{k'}, B_{k'})\), based on channel capacity \(C\) and buffer level \(B\). FD is subject to residual self-interference (RSI).    
    \item \textbf{FlexD:} Transmission occurs from the user with higher \(\min(C, B)\), i.e., greater effective transmission potential. Only one user transmits per slot. If the selected user's buffer is empty, the system defaults to HD-like behavior.
\end{itemize}

\begin{table}[t]
    \centering
    \small 
    \begin{tabular}{@{} l p{3cm} p{4cm} @{}}
        \toprule
        \textbf{Mode} & \textbf{Transmission Direction} & \textbf{Scheduling Strategy}  \\
        \midrule
        HD & One user at a time (buffer-based) & Fixed random direction or buffer availability \\ \midrule
        FD & Both users simultaneously (if data exists) & Buffer-aware bidirectional \\ \midrule
        FlexD & One user at a time (channel- and buffer-aware) & Buffer-aware and opportunistic direction \\
        \bottomrule
    \end{tabular}
    \caption{Comparison of HD, FD, and FlexD traffic models}
    \label{table:1}\vspace{-5mm}
\end{table}

Fig.~\ref{fig:que} presents the throughput outage vs target rate threshold \(R_t\) for FlexD, HD, and FD (with fixed RSI power \(\sigma^2_{\text{rsi}} = -150\) dBm), for a user pair \((\textsf{U}_k, \textsf{U}_{k'})\) assisted by a RIS with \(M = 128\). Packet arrivals follow a Poisson process with mean rate $0.8$. FlexD consistently outperforms HD and FD, benefiting from its adaptive direction selection based on instantaneous channel and buffer states. In contrast, HD and FD exhibit higher outage at lower thresholds due to their fixed, non-adaptive transmission strategies. The results highlight FlexD’s improved buffer utilization and reduced queuing delay, underscoring its suitability for delay-sensitive applications such as vehicular and mission-critical networks. This baseline model evaluates FlexD, HD, and FD performance, with future work extending to more realistic models incorporating MAC layer dynamics, adaptive scheduling, and interference management.

\vspace{-0.2cm}
\section{Conclusion}
\vspace{-0.2cm}
This paper investigates the integration of Reconfigurable Intelligent Surfaces (RIS) with Flexible-Duplex (FlexD) communication to enhance system throughput in multi-user cooperative networks by mitigating external interference. We propose a user-pair selection scheme leveraging RIS reconfigurability and FlexD’s dynamic scheduling, even in the absence of full CSI of the interferer. Closed-form expressions for throughput outage probability are derived, with asymptotic analysis demonstrating FlexD’s superiority over conventional HD and FD systems under interference. Numerical results validate the proposed scheme, highlighting its potential for data management and energy efficiency. These findings highlight RIS-assisted FlexD's potential to enhance wireless network performance in interference-prone environments, with applications in next-generation networks. Future work will explore distributed MAC protocols for dynamic scheduling and channel estimation~\cite{Zhang_2025,Zhang_optimal}.

\section*{APPENDIX:Proof of \eqref{outage:case1}}

\label{appendix:A}
To find the outage probability described in \eqref{outage}, we must identify each RV in $\gamma_{k,k'}$ (i.e., same RVs in $\gamma_{k',k}$). Consider $Z_1=(\sum_{m=1}^{M}\alpha_{k,m}\alpha_{m,k'}),Z_2=(\sum_{m=1}^{M}\beta_{m}{\alpha_{m,k'}} e^{-j(\phi_m-\theta_{k,m})})$. In the next section, we discuss the behavior of each RV in $\gamma_{k,k' }$ and the dependencies between them since, $\alpha_{m,k'}$ is common for both RVs. We can approximate the $Z_1$ with complex Gaussian RV for large $M$, since it reduces the complexity when deriving a distribution for the sum of the product of two Rayleigh RVs. The expectation and variance of $Z_1$ denoted as $\mu, s$ respectively, can be computed as $\mu = \sum_{m=1}^{M} \mathbb{E}[\alpha_{k,m}]\mathbb{E}[\alpha_{m,k'}] = \sum_{m=1}^{M}\left(\frac{\sigma}{\sqrt{2}}\frac{\sqrt{\pi}}{\sqrt{2}}\right)^2\label{mean_z1} , s = \mathbb {E}[(Z_1 - \mathbb {E}[Z_1])^{2}]=\mathbb {E}\left [{\left ({\sum \limits _{m=1}^{M}\left ({\alpha_{k,m} \alpha_{m,k'} - \mathbb {E}[\alpha_{k,m} \alpha_{m,k'}]}\right)}\right)^{2}}\right] $ \mbox{$= \sum \limits _{i=1}^{M} \sum \limits _{j=1}^{M}\left ({\mathbb {E}\left [{\alpha_{k,i} \alpha_{k,j}}\right] \mathbb {E}\left [{\alpha_{i,k'} \alpha_{j,k'}}\right]}\right) - \frac {\sigma^4\pi ^{2}}{16} M^{2} \label{variance_z1}.\notag
$}
Mathematical relationship between $\mathbb {E}[|h_{k,i}||h_{k,j}|]$ and $\mathbb {E}[h_{k,i}h^*_{k,j}]$ is described as the $(i,j)$ entry in matrices $\bar{\mathbf{\Omega}}$ and $\mathbf{\Omega}$ into $\sigma^2$ respectively \cite{coon_com_let_2022}. Therefore, we have expectation \({\mathbb{E}[\alpha_{k,i} \alpha_{k,j}] = \mathbb{E}[\alpha_{i,k'} \alpha_{j,k'}] = \sigma^2 [\bar{\mathbf{\Omega}}]_{i,j}}\).
 The expectation of \(Z_2\) can be computed directly as follows here: \mbox{\(\mathbb{E}[Z_2] = \sum_{m=1}^{M} \beta_m \alpha_{m,k'} e^{-j(\phi_m - \theta_{k,m})} = 0\)}, where phase term \mbox{\(\phi_m - \theta_{k,m} \in (-\pi, \pi)\)}. Likewise, variance \({\mathrm{Var}}[Z_2] = \mathbb{E}[(Z_2 - \mathbb{E}[Z_2])^2]\) can be found. 
\\Next, we focus on the covariance between \(Z_1, Z_2\). \\\(\mathrm{Cov}(Z_1, Z_2) \hspace{-1mm}= \hspace{-1mm}\mathbb{E}\bigg[\sum_{m,n=1}^{M} \alpha_{k,m} \alpha_{m,k'} \alpha_{n,k'} \beta_{n} e^{-(\phi_n - \theta_{k,n})}\bigg] \hspace{-1mm} - \hspace{-1mm} \sum_{m=1}^{M} \mathbb{E}[\alpha_{k,m} \alpha_{m,k'}] \sum_{n=1}^{M} \mathbb{E}[\alpha_{n,k'} \beta_n e^{-(\phi_n - \theta_{k,n})}] = 0\).

Therefore, \(Z_1\) and \( Z_2\) are uncorrelated. 
For large \(M\), \(X_1 = Z_1^2, Y_1 = |Z_2|^2 \sim \exp(\rho)\), where \(\rho = 1 / \mathrm{Var}[Z_2]\). Hence, the PDFs of $X_1$ and $Y_1$ are as follows: 
\begin{align*}
    &f_{X_1}(x_1) = (e^{-\left(\sqrt{x_1}-\mu\right)^2/{2s}}+e^{-\left(\sqrt{x_1}+\mu\right)^2/{2s}})/2\sqrt{2x_1\pi s}\\& f_{Y_1}(y_1) = \rho e^{-\rho y_1}
\end{align*}
Due to the intractability of  the outage probability in \eqref{outage}, we take the expected values of each RV in the direction selection part in which $\mathbb{E}[X_1]$ and $M\sigma^2$, where $M\sigma^2$ is the expected value of the Gaussian approximated $\sum_{m=1}^{M}{\alpha_{m,k'}^2}$. Hence, the \eqref{outage} will reduce to the following form. $F_{\gamma_{{k  \leftrightarrow k'}}}\left(\tau\right) = \Pr\left({\gamma}_{k,k'} < \tau \ \& \  \quad
 \eta_{I,k} \geq \eta_{I,k'} \right) + \Pr\left({\gamma}_{k,k'} < \tau \ \ \&  \quad
  \eta_{I,k} < \eta_{I,k'}\right)\notag
$. For $\eta_{I,k} \geq \eta_{I,k'}$
\begin{align*}
    F_{\gamma_{{k  \leftrightarrow k'}}}\left(\tau\right)=& 
    \Pr\left(\frac{\bar{\gamma}_{k,k'} X_1}{\eta_{I,k'}Y_1+1} < \tau\right)\\=&1- \int_{\frac{\tau}{\bar{\gamma}_{k,k'}}}^{\infty} F_{Y_{1}}\left(\frac{\bar{\gamma}_{k,k'} x_1 -\tau}{\tau \eta_{I,k'}}\right)f_{X_{1}}(x_{1}) \, d{x_{1}}
\end{align*} 
The final expression is obtained by substituting $F_{Y_{1}}(y)$ and $f_{X_{1}}(x)$, followed by integration. Where $F_{Y_{1}}(y)=1-e^{-\rho y_1}$ is the CDF of $Y_1$. Similarly, for $\eta_{I,k} < \eta_{I,k'}$, $F_{\gamma_{{k  \leftrightarrow k'}}}\left(\tau\right)$ can be obtained.


\end{document}